\documentclass[12pt,preprint]{aastex}

\usepackage{rotating}
\shorttitle{}  
\shortauthors{Katsuda et al.}

\begin{document}

\title{Spatially-Resolved Spectroscopy of a Balmer-Dominated Shock in the Cygnus Loop: An Extremely Thin Cosmic-Ray Precursor?}

\author{Satoru Katsuda\altaffilmark{1}, 
Keiichi Maeda\altaffilmark{2,3},
Yutaka Ohira\altaffilmark{4},
Yoichi Yatsu\altaffilmark{5},
Koji Mori\altaffilmark{6},
Wako Aoki\altaffilmark{7},
Kumiko Morihana\altaffilmark{8},
John C. Raymond\altaffilmark{9},
Parviz Ghavamian\altaffilmark{10},
Jae-Joon Lee\altaffilmark{11},\\
Jiro Shimoda\altaffilmark{4},
Ryo Yamazaki\altaffilmark{4}
}

\altaffiltext{1}{Institute of Space and Astronautical Science (ISAS), Japan Aerospace Exploration Agency (JAXA), 3-1-1 Yoshinodai, Chuo, Sagamihara, Kanagawa 252-5210, Japan}

\altaffiltext{2}{Department of Astronomy, Kyoto University, Kitashirakawa-Oiwake-cho, Sakyo-ku, Kyoto 606-8502, Japan}

\altaffiltext{3}{Kavli Institute for the Physics and Mathematics of the Universe (WPI), The University of Tokyo, 5-1-5 Kashiwanoha, Kashiwa, Chiba 277-8583, Japan}

\altaffiltext{4}{Department of Physics and Mathematics, Aoyama Gakuin University, 5-10-1 Fuchinobe, Sagamihara 252-5258, Japan}

\altaffiltext{5}{Department of Physics, Tokyo Institute of Technology, 2-12-1 Ohokayama, Meguro, Tokyo 152-8551, Japan}

\altaffiltext{6}{Department of Applied Physics and Electronic Engineering, Faculty of Engineering, University of Miyazaki, 1-1 Gakuen Kibanadai-Nishi, Miyazaki 889-2192, Japan}

\altaffiltext{7}{National Astronomical Observatory of Japan, Mitaka, Tokyo 181-8588, Japan}

\altaffiltext{8}{Nishi-Harima Astronomical Observatory, Center for Astronomy, University of Hyogo, 407-2 Nichigaichi, Sayo-cho, Sayo, Hyogo, 670-5313}

\altaffiltext{9}{Harvard-Smithsonian Center for Astrophysics, 60 Garden St., Cambridge, MA 02138, USA}

\altaffiltext{10}{Department of Physics, Astronomy and Geosciences, Towson University, Towson, MD 21252, USA}

\altaffiltext{11}{Korea Astronomy and Space Science Institute, Daejeon 305-348, Korea 0000-0003-0894-7824}

\begin{abstract}
We present high-resolution long-slit spectroscopy of a Balmer-dominated shock in the northeastern limb of the Cygnus Loop with the {\it Subaru} high dispersion spectrograph.  By setting the slit angle along the shock normal, we investigate variations of the flux and profile of the H$\alpha$ line from preshock to postshock regions with a spatial resolution of $\sim$4$\times10^{15}$\,cm.  The H$\alpha$ line profile can be represented by a narrow (28.9$\pm$0.7\,km\,s$^{-1}$) Gaussian in a diffuse region ahead of the shock, i.e., a photoionization precursor, and narrow (33.1$\pm$0.2\,km\,s$^{-1}$) plus broad (130--230\,km\,s$^{-1}$) Gaussians at the shock itself.  We find that the width of the narrow component abruptly increases up to 33.1$\pm$0.2\,km\,s$^{-1}$, or 38.8$\pm$0.4\,km\,s$^{-1}$ if we eliminate projected emission originating from the photoionization precursor, in an unresolved thin layer ($\lesssim4\times10^{15}$\,cm at a distance of 540\,pc) at the shock.  We show that the sudden broadening can be best explained by heating via damping of Alfv\'en waves in a thin cosmic-ray precursor, although other possibilities are not fully ruled out.  The thickness of the cosmic-ray precursor in the Cygnus Loop (a soft gamma-ray emitter) is an order of magnitude thinner than that in Tycho's Knot g (a hard gamma-ray emitter), which may be caused by different energy distribution of accelerated particles between the two sources.  In this context, systematic studies might reveal a positive correlation between the thickness of the cosmic-ray precursor and the hardness of the cosmic-ray energy distribution.  
\end{abstract}
\keywords{acceleration of particles --- ISM: individual objects (Cygnus Loop) --- ISM: supernova remnants --- shock waves}

\section{Introduction}

Supernova explosions produce strong shock waves propagating into the interstellar medium (ISM).  Whereas these shocks are usually associated with bright optical forbidden lines from metals (radiative shocks) and permitted lines from hydrogen, a small subsample of supernova remnants (SNRs) exhibit faint Balmer-dominated shocks which are characterized by H lines and very weak forbidden lines from lowly-ionized metals \citep[e.g.,][for reviews]{2001SSRv...99..209R,2010PASA...27...23H,2013SSRv..178..633G}.  Diagnostics of the H line profile have been used to probe ``collisionless" shocks, where the shock transition occurs effectively by collective interactions of the plasma with the magnetic field rather than Coulomb collisions.  The H line profile behind a Balmer-dominated shock consists of narrow and broad components.  The narrow component arises from collisional excitation of cold neutrals that pass through the shock, and the broad component arises from collisional excitation and cascading processes of hot neutrals that are created by charge exchange (CX) reactions between hot downstream protons heated at the shock and the cold neutrals unaffected by the shock transition region \citep{1980ApJ...235..186C,2001ApJ...547..995G}.  Therefore, the widths of the narrow and broad components should reflect temperatures of unshocked neutrals and shocked protons, respectively.  

It has been a long-standing problem that the widths of the narrow component, FWHMs $\sim30-50$\,km\,s$^{-1}$, are broader than the thermal width expected in the ISM, roughly 21 ($T$/10$^4$\,K)$^{0.5}$\,km\,s\,$^{-1}$ \citep[e.g.,][]{2003A&A...407..249S}.  Although the broadening can be partly due to heating by a photoionization precursor from postshock emission such as He II $\lambda$304 \citep[e.g.,][]{2000ApJ...535..266G}, this effect is insufficient to quantitatively explain the width observed.  Various possibilities have been considered in the interpretation of the nature of the additional broadening \citep{1994ApJ...420..721H,1994ApJ...420..286S}.  Of these, the most likely is a cosmic-ray (CR) precursor, in which the gas is heated by damping of Alfv\'en/sound waves emitted by CRs and/or is disturbed by Alfv\'en wave turbulence.  \citet{2007ApJ...659L.133L,2010ApJ...715L.146L} discovered a possible CR precursor associated with a Balmer-dominated shock at the eastern edge \citep[Knot g:][]{1978ApJ...224..851K} in Tycho's SNR, where both the intensity and width of the H$\alpha$ line gradually increases within a thin ($\sim$10$^{16}$\,cm) region ahead of the shock.  \citet{2009ApJ...690.1412W} showed that the precursor can heat the gas by damping of sound waves from CRs via an acoustic instability.  So far, it is the only known example of a promising CR precursor \citep[see also][reporting a hint of the CR precursor from Kepler's SNR]{2003A&A...407..249S}.  Therefore, it is essential to accumulate observational information about CR precursors by observing other Balmer-dominated shocks. 

Balmer-dominated shocks in the Cygnus Loop offer a unique opportunity for the study of thin precursors, thanks to its proximity --- it is the closest \citep[$d=$540$^{+100}_{-80}$\,pc:][]{2005AJ....129.2268B} among 11 SNRs (both in our Galaxy and the LMC) associated with Balmer-dominated shocks, with 1$^{\prime\prime}$ corresponding to $2.6\times10^{-3}$\,pc.  Recently, \citet{2014ApJ...791...30M} performed high-resolution spectroscopy of a number of shocks along with diffuse emission 2.5$^{\prime}$ ahead of northeastern (NE) shocks of the Cygnus Loop.  They found that H$\alpha$ line widths in the diffuse regions are as narrow as $\sim$29\,km\,s$^{-1}$, while H$\alpha$ profiles of the shocks themselves consist of broader narrow ($\sim$36\,km\,s$^{-1}$) and broad ($\sim$250\,km\,s$^{-1}$) components.  They suggested that the H$\alpha$ broadening of $\sim$29\,km\,s$^{-1}$ in the diffuse region is due to photoionization, and speculated that the additional broadening of the narrow component at the shock originates from a thin CR precursor.  Like Tycho's Knot g \citep{2007ApJ...659L.133L}, such a thin precursor may be detected by spatially-resolved spectroscopy of the Balmer-dominated shock.

Here, we present high-resolution long-slit spectra which are spatially resolved perpendicular to the best-studied Balmer-dominated shock in the Cygnus Loop \citep{1983ApJ...275..636R,1985ApJ...295...43F,1992ApJ...400..214L,1994ApJ...420..721H,2000AJ....120.1925S,2005AJ....129.2268B,2008ApJ...680.1198K,2014ApJ...791...30M}.  Consistent with the previous optical spectroscopy \citep{2014ApJ...791...30M}, we detect a diffuse precursor ahead of the shock throughout the slit position.  We find that the width of the narrow component increases within an unresolved region at the shock, which we attribute to heating by damping of Alfv\'en waves in a CR precursor.  We present our observations and results in Section~2, and interpretations in Section~3.

\section{Observations and Results}

\begin{figure}
\begin{center}
\includegraphics[angle=0,scale=0.55]{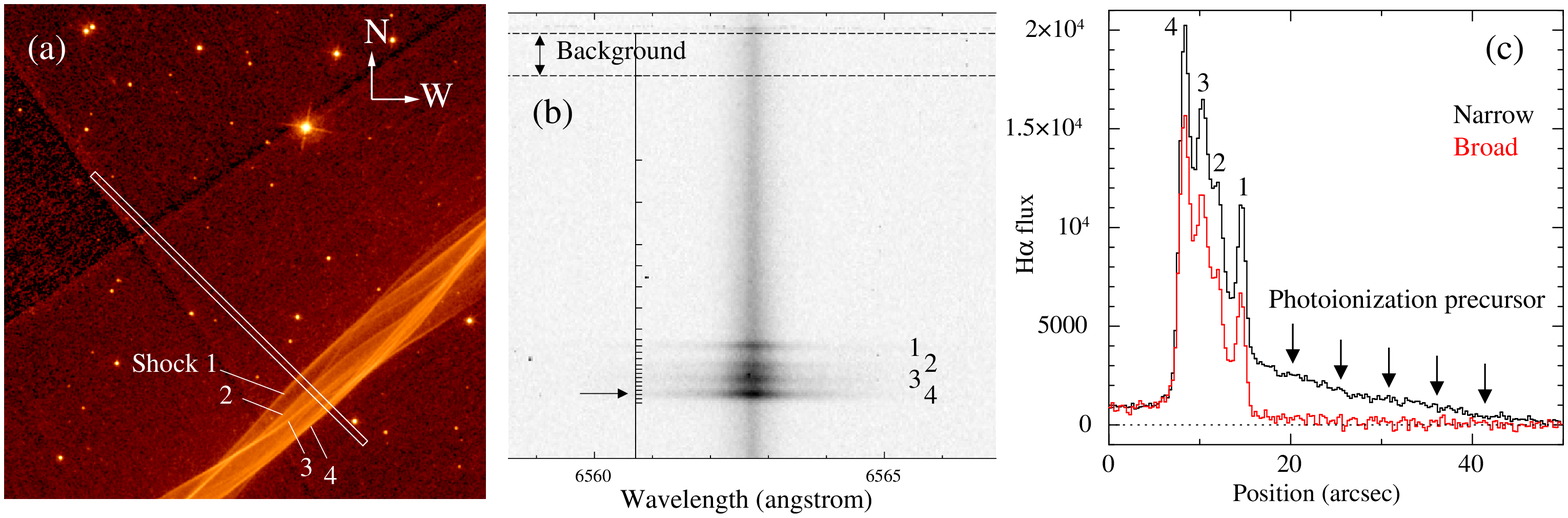}\hspace{1cm}
\caption{(a) H$\alpha$ image of the Balmer-dominated shock in the Cygnus Loop obtained by {\it HST} WFPC2 camera \citep{2005AJ....129.2268B}.  The 1$^{\prime\prime} \times 60^{\prime\prime}$ slit of the {\it Subaru} HDS is overlaid as a white rectangle.  (b) H$\alpha$ two-dimensional spectrum with the {\it Subaru} HDS.  The vertical scale bar indicates spectral extraction regions, among which the top region is used as our background.  The spectrum in Fig.~\ref{fig:spec} is extracted from the region marked by a horizontal arrow.  (c) Background-subtracted H$\alpha$ profiles along the slit position.  The narrow and broad components are taken from velocity ranges of $|v_{\rm LSR}| <$ 25\,km\,s$^{-1}$ and $25$\,km\,s$^{-1}$ $< |v_{\rm LSR}| <$ 125\,km\,s$^{-1}$, respectively.
} 
\label{fig:image}
\end{center}
\end{figure}

We observed one of the brightest Balmer-dominated shocks in the NE limb of the Cygnus Loop on 2015 August 31, using the {\it Subaru} High Dispersion Spectrogram \citep[HDS:][]{2002PASJ...54..855N}.  We use a long slit together with the H$\alpha$ order-blocking filter.  The slit width is 1$^{\prime\prime}$, which gives velocity resolution of FWHM=9\,km\,s$^{-1}$.  The slit was centered at (RA, Decl.) = (20:56:04.8, +31:56:46.7) (J2000), with a position angle of 46$^{\circ}$ measured north of east as shown in Fig.~\ref{fig:image} (a).   This way, we investigate spatial variations of the line profile along the shock normal, i.e., parallel to the shock motion.  The total exposure time was 6$\times$30 minutes, which was reduced to 2.5 hr after rejecting one frame affected by fog.  The spectrum was binned by 2 pixels for both the dispersion axis and the slit direction, so that the pixel scales after the binning were 0.0364\AA/pixel and 0.276$^{\prime\prime}$/pixel, respectively.  The wavelength range covered by this observation is 6515--6589\AA.  The seeing was about 0.5$^{\prime\prime}$.  We performed a standard processing of the {\it Subaru} data, including overscan correction, flat fielding, and wavelength calibration based on the spectrum from a Th-Ar lamp.  We reject cosmic-ray backgrounds, by taking the median of the four frames.  In this processing, we utilize the version 2.16 of the IRAF software \footnote{http://iraf.noao.edu/}.

Figure~\ref{fig:image} (b) shows a fully processed two-dimensional spectrum of the H$\alpha$ line, for which the x- and y-axis are responsible for the wavelength and the slit position (top is to the NE), respectively.  The four blobs near the bottom correspond to the Balmer-dominated shocks 1--4.  Since we did not obtain a sky background frame separately, we adopt the far upstream region between two dotted horizontal lines in Fig.~\ref{fig:image} (b) as our background.  The background-subtracted H$\alpha$ intensity profiles along the slit position are plotted in Fig.~\ref{fig:image} (c), where the black and red are responsible for narrow and broad components, respectively.  The broad component arises immediately behind the bright shocks, whereas the narrow component extends ahead of them as indicated by arrows with ``photoionization precursor" in the plot (see below).  

\begin{figure}
\begin{center}
\includegraphics[angle=0,scale=0.6]{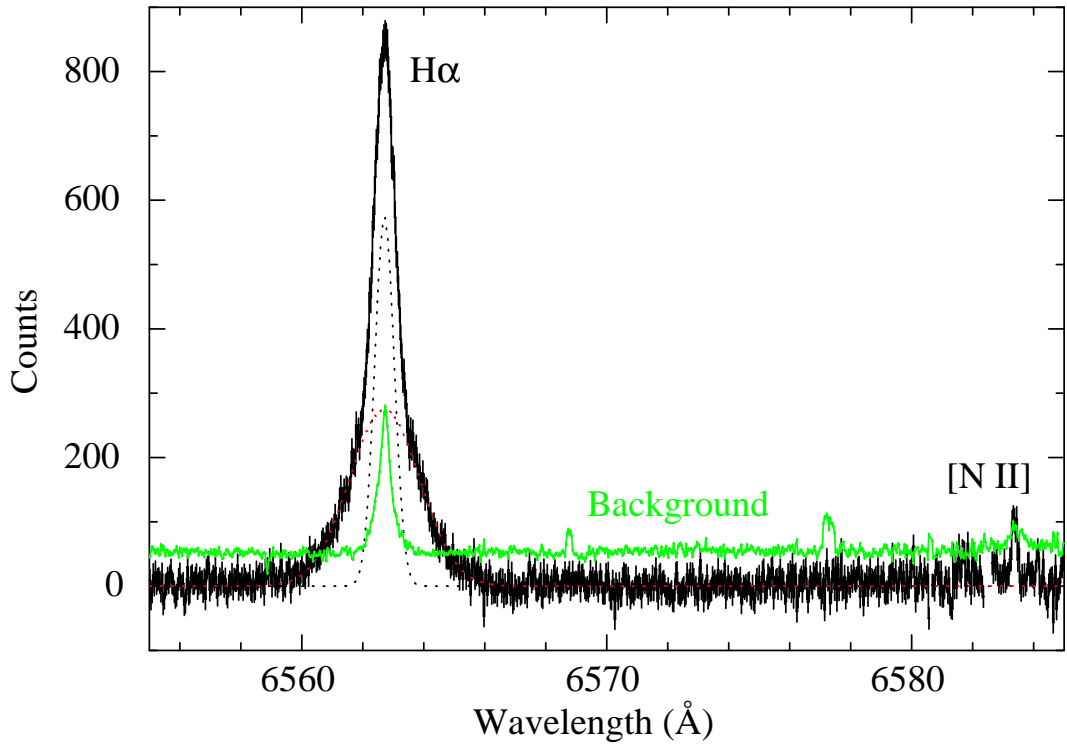}\hspace{1cm}
\caption{Background-subtracted {\it Subaru} HDS spectrum from the brightest region marked in Fig.~\ref{fig:image} (b).  The H$\alpha$ profile is fitted by narrow (black dotted line) and broad (red dotted line) Gaussians.  The background spectrum is plotted in green.
} 
\label{fig:spec}
\end{center}
\end{figure}

To investigate spatial variations of the H$\alpha$ profile, we extract one-dimensional spectra from 21 regions along the slit, as indicated by a vertical scale bar in Fig.~\ref{fig:image} (b).  The regions are divided such that brighter regions have smaller widths with a minimum width of 2 pixels or 0.55$^{\prime\prime}$.  Figure~\ref{fig:spec} shows a background-subtracted one-dimensional spectrum taken from the brightest region indicated by a horizontal arrow in Fig.~\ref{fig:image} (b), together with the background spectrum in green.  The H$\alpha$ line profile can be reproduced by two Gaussian components: a broad component (dotted red line) arising from a shock-heated gas and a narrow component (dotted black line) arising from unshocked neutrals.  Another small peak at 6583.4\AA\ is the [N II] line originating from the Cygnus Loop itself.  Below, we will fit both of these two lines by either a single Gaussian or double Gaussians, after taking account of instrumental broadening of FWHM=9\,km\,s$^{-1}$. 

Our analysis shows that only a narrow Gaussian component is required in the precursor, with an upper limit of broad-to-narrow intensity ratios ($I_{\rm b}/I_{\rm n}$) of 0.3.  On the other hand, two Gaussian components are required at the Balmer-dominated shocks and its interior where $I_{\rm b}/I_{\rm n}$ ratios range from 0.75$\pm$0.02 to 2.01$\sim$0.08.  Such a strong $I_{\rm b}/I_{\rm n}$ difference between the precursor and the Balmer-dominated shock clearly demonstrates that the precursor is not the result of geometric projection of fainter Balmer-dominated shocks, but an intrinsic precursor associated with the bright shocks.  

\begin{figure}
\begin{center}
\includegraphics[angle=0,scale=0.6]{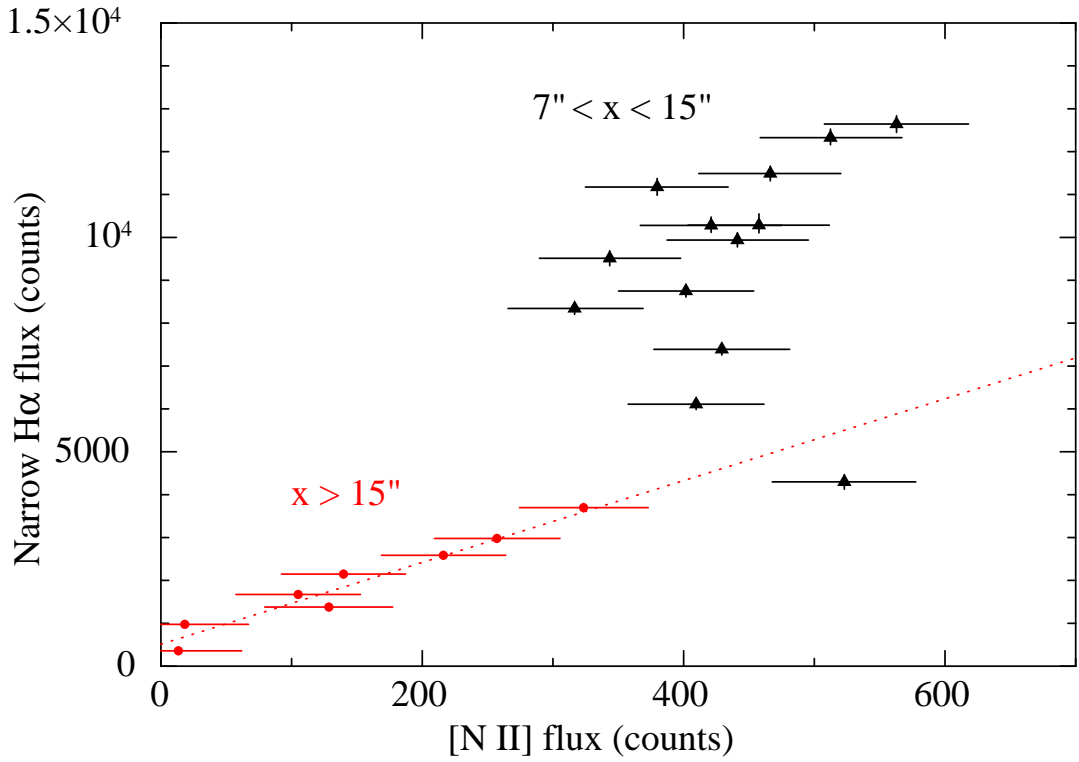}\hspace{1cm}
\caption{Flux of the narrow H$\alpha$ line vs.\ flux of the [N II] line.  The red data (filled circles) are from the photoionization precursor ($x > 15^{\prime\prime}$; cf. Fig.~\ref{fig:results}), while the black data (filled triangles) are from the Balmer-dominated shocks ($7^{\prime\prime} < x < 15^{\prime\prime}$).  As indicated by a red dotted line, there is a linear correlation between the H$\alpha$ intensity and the [N II] intensity in the photoionization precursor.
} 
\label{fig:nHa_vs_N2}
\end{center}
\end{figure}

The [N II] line profiles can be fitted by a single, narrow Gaussian throughout the slit.  We measure the width of the [N II] line to be narrower than FWHM=8\,km\,s$^{-1}$ (after considering the instrumental resolution) all along the slit position, which allows us to infer the N temperature to be $T<20,000$\,K, consistent with the H temperature in the photoionization precursor.  Also, there is a good correlation between the H$\alpha$ flux and [N II] flux within the photoionization precursor ($I_{{\rm H}{\alpha}} = 9.37 \times I_{\rm [N II]} + 511$ photons), as shown in Fig.~\ref{fig:nHa_vs_N2}.  These facts lead us to conclude that the [N II] line entirely arises from the photoionization precursor.  This idea is in agreement with the earlier interpretation for the origin of the extremely faint forbidden line emission including [N II] and [S II] lines from the relevant Balmer-dominated shock \citep{1985ApJ...295...43F}.  

\begin{figure}
\begin{center}
\includegraphics[angle=0,scale=0.8]{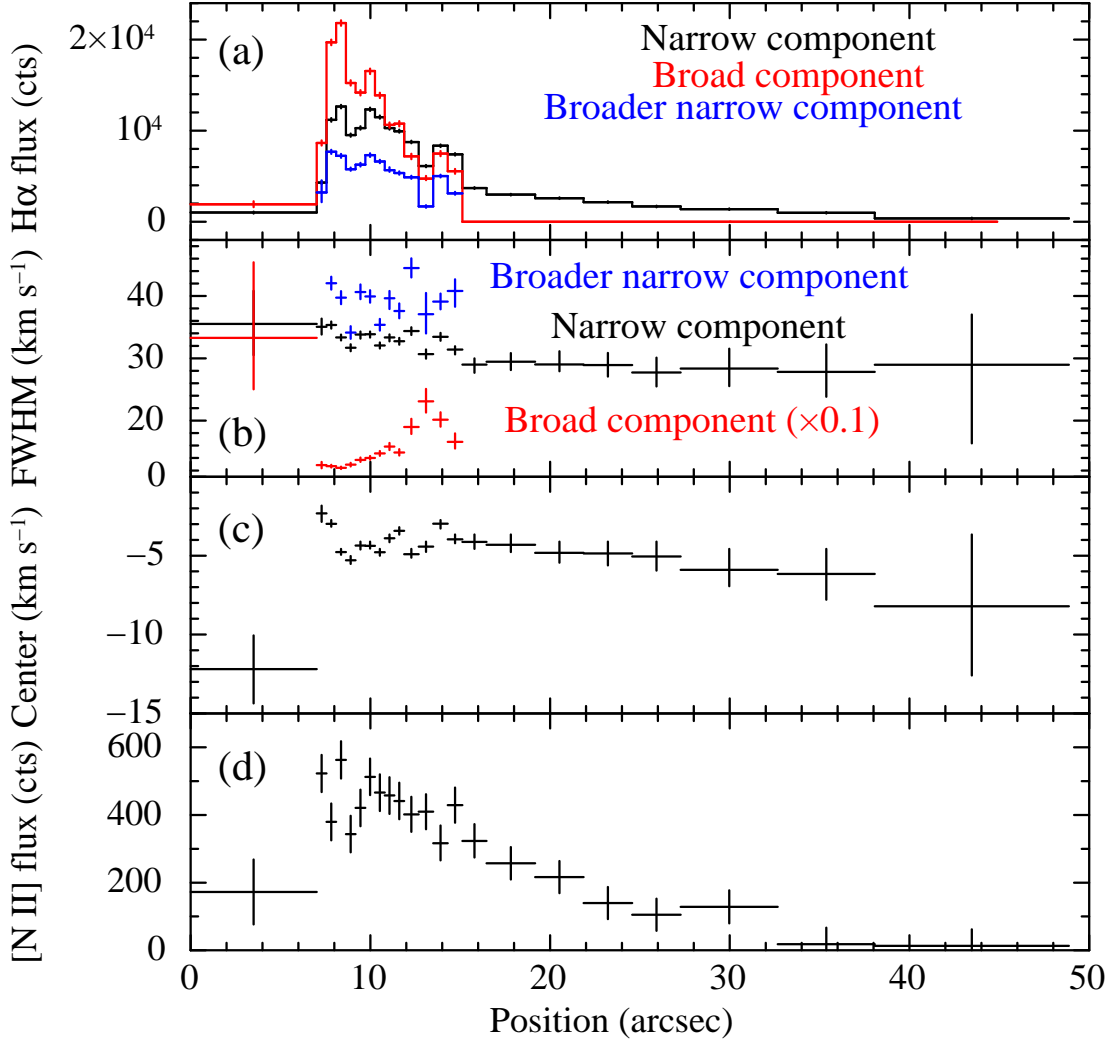}\hspace{1cm}
\caption{(a) H$\alpha$ fluxes along the slit position.  Black, red, and blue are responsible for the narrow, broad, and broader-narrow components, respectively (see the text for explanations).  The errors represent 1-$\sigma$ statistical uncertainties.  (b) Same as top but for FWHMs.  Values for the broad component are multiplied by 0.1.  (c) Same as top but for line centers.  The centers are taken to be the same between the narrow and broad components.  (d) Same as top but for [N II], for which only the narrow component is detected.
} 
\label{fig:results}
\end{center}
\end{figure}

Figure~\ref{fig:results} shows the best-fit Gaussian parameters as a function of the slit position, where $x=0^{\prime\prime}$ corresponds to the bottom line of the two-dimensional spectrum in Fig.~\ref{fig:image} (b).  Within the precursor region ($x > 15^{\prime\prime}$), the width of the H$\alpha$ line and the H$\alpha$/[N II] intensity ratio are roughly constant at 28.9$\pm$0.7\,km\,s$^{-1}$ (error-weighted mean FWHM) and 12.1$\pm$0.2, respectively, fully consistent with those of a photoionization precursor in the NE Cygnus Loop \citep{2014ApJ...791...30M}.  We find from Fig.~\ref{fig:results} (b) that the narrow component's width abruptly increases from the precursor region to 33.1$\pm$0.2\,km\,s$^{-1}$ (error-weighted mean FWHM) within a remarkably thin (unresolved) layer at the shock position.  We have ruled out a possibility that the sudden broadening is an artifact due to the broad component, based on the fact that the line width does not change much whether or not we introduce the broad component in the narrow-band fitting.  The flux of the narrow component in the photoionization precursor gradually increases from a far upstream region to the shock front.  This may be caused by projection effects --- increasing line-of-sight column densities toward the shock, which would be applicable to the relevant shock showing an interaction with a cloud \citep{1992ApJ...400..214L,1994ApJ...420..721H}.  The line centroids of the broad and narrow components, which are taken to be the same between the narrow and broad components in our analysis, do not show significant spatial variations except for the innermost region ($x < 7^{\prime\prime}$).  

As for the broad component, the width decreases from the outermost shock (230$\pm$20\,km\,s$^{-1}$) to the innermost shock (130$\pm$5\,km\,s$^{-1}$) within the four bright Balmer-dominated shocks.  The difference by a factor of $\sim$2 can be reasonably interpreted as a result of different shock speeds; we have checked that the outermost Balmer-dominated shock is moving $\sim$1.5 times faster than the other three shocks, based on our own proper-motion measurements by data from the {\it Hubble Space Telescope} ({\it HST}).  We see an even broader (330$^{+120}_{-80}$\,km\,s$^{-1}$) line at the innermost region.  Some of the width may be the projected bulk speed of the shocked gas, if the region is projected at a significant angle, and if both front and back surfaces contribute.

\begin{figure}
\begin{center}
\includegraphics[angle=0,scale=0.6]{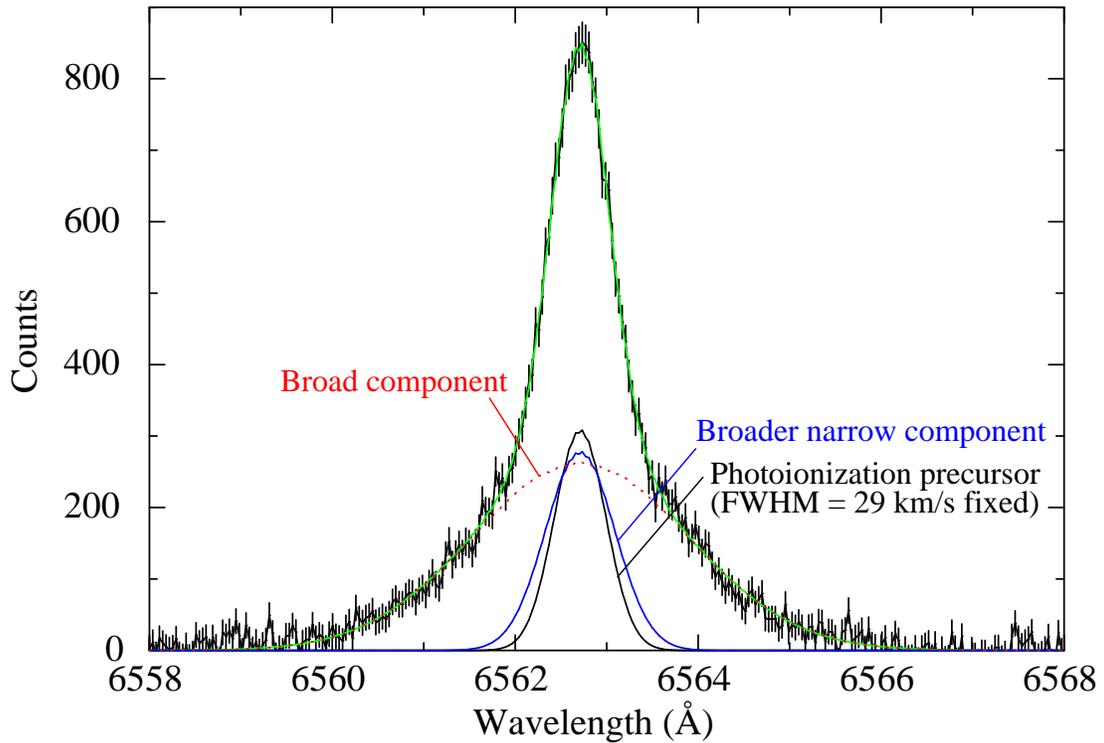}\hspace{1cm}
\caption{Same spectrum in Fig.~\ref{fig:spec}, but the H$\alpha$ profile is fitted by three Gaussians, i.e., narrow (black solid line), broader narrow (blue solid line), and broad (red dotted line) components.  
} 
\label{fig:spec_3gau}
\end{center}
\end{figure}

The presence of the [N II] line at the bright Balmer-dominated shocks (x$<15^{\prime\prime}$) suggests a possible contamination from the photoionization precursor due to projection.  In other words, the H$\alpha$ line at the Balmer-dominated shock is a mixture of the narrow (29\,km\,s$^{-1}$) and a broader narrow ($>$33\,km\,s$^{-1}$) component.  Other hints of such a contamination include (1) the $I_{\rm b}/I_{\rm n}$ ratio gradually decreases from the innermost shock to the outermost shock (Fig.~\ref{fig:results} (a)), and (2) the narrow component's width becomes narrower at inter-blob regions than blobs themselves (Fig.~\ref{fig:results} (a) and (b)).  Therefore, we try to eliminate the contamination from the photoionization precursor.  Using the [N II] flux and the relation between $I_{{\rm H}{\alpha}}$ and $I_{\rm [N II]}$, we estimate the contaminating H$\alpha$ flux.  We then refit the H$\alpha$ profiles with three Gaussian components responsible for narrow (photoionization precursor), broader narrow (thin precursor), and broad (postshock) components, as shown in Fig.~\ref{fig:spec_3gau}.  In this way, we find out broader-narrow components (the blue data in Fig.~\ref{fig:results} (a)), whose error-weighted mean FWHM is 38.8$\pm$0.4\,km\,s$^{-1}$ (the blue data in Fig.~\ref{fig:results} (b)). 

\section{Summary and Discussions}

For a Balmer-dominated shock in the NE Cygnus Loop, we have found that the narrow H$\alpha$ component abruptly broadens from 28.9$\pm$0.7\,km\,s$^{-1}$ up to 38.8$\pm$0.4\,km\,s$^{-1}$ (after removing the contamination from a photoionization precursor) within an extremely thin layer ($\lesssim$0.0013\,pc) at the shock position.  Here, we assess the nature of this abrupt broadening.  As discussed in \citet{1994ApJ...420..286S}, there are no promising postshock mechanisms (including the conversion of Ly$\beta$ $\to$ H$\alpha$, momentum transfer in the process of excitation, elastic collisions between narrow component neutrals and shock heated protons, and molecular dissociation), leaving only preshock mechanisms.  Of these, there are two major possibilities: (1) a fast neutral precursor, in which fast neutrals returning from the postshock heat the preshock gas, and (2) a CR precursor described in Section~1.  

The fast neutral scenario was later disfavored by a simulation by \citet{1996MNRAS.280..103L}.  The authors found that the thickness of the neutral precursor becomes only $\sim$10$^{14} \left(\frac{n}{1\,{\rm cm}^{-3}}\right)^{-1}$\,cm, which is smaller than the characteristic length ($\sim$3$\times$10$^{14} \left(\frac{n}{1\,{\rm cm}^{-3}}\right)^{-1}$\,cm) for CX between upstream protons and neutrals, violating a critical condition that at least one CX reaction must occur to generate warm neutrals; upstream (unshocked) protons are first heated, and then these warm protons will become warm neutrals via CX with upstream cold neutrals.  These characteristic lengths can be also checked by analytical solutions \citep[Equations 15 and 17 in][]{2012ApJ...758...97O} combined with realistic physical parameters for the Cygnus Loop.  In addition, \citet{2012ApJ...760..137M}, \citet{2013PhRvL.111x5002O}, and \citet{2016ApJ...817..137O} also confirmed that the returning fast neutrals do not affect the width of the narrow H$\alpha$ line for shock velocities $\lesssim$1500\,km\,s$^{-1}$ like those in the Cygnus Loop.  Meanwhile, our measured upper limit of the precursor's length ($\lesssim$4$\times10^{15}$\,cm) does not enable us to eliminate (nor support) the possibility of the heating by returning fast neutrals.  Thus, precise determination of the precursor size with better spatial resolution is desired to clarify whether or not this idea is still viable.  

As for the CR precursor scenario, there are two ways to broaden the line.  One is Doppler broadening due to Alfv\'en or magneto sonic turbulence, and the other is the heating via damping of these turbulence.  In any case, the affected (warm) protons need to become neutrals via CX with upstream cold neutrals.  This condition sets a lower limit of the CR precursor size at the characteristic CX length scale of a few 10$^{14}$\,cm, which is met for the precursor of our interest ($L_{\rm cr}\lesssim4\times10^{15}$\,cm).  

We now examine if the two waves can really grow in the thin CR precursors.  The time scales for Alfv\'en waves and sound waves to grow are 
\[
t_{\rm a} \sim 5\times10^{7}\,{\rm s} \left(\frac{V_{\rm a}}{6.6\,{\rm km}\,{\rm s}^{-1}}\right) \left(\frac{L_{\rm cr}}{0.001\,{\rm pc}}\right) \left(\frac{V_{\rm sh}}{200\,{\rm km}\,{\rm s}^{-1}}\right)^{-1} \left(\frac{\eta_{\rm cr}}{0.1}\right)^{-1}
\] \citep{1978MNRAS.182..147B}
\[
t_{\rm s} \sim 5\times10^{8}\,{\rm s} \left(\frac{V_{\rm so}}{30\,{\rm km}\,{\rm s}^{-1}}\right) \left(\frac{L_{\rm cr}}{0.001\,{\rm pc}}\right) \left(\frac{V_{\rm sh}}{200\,{\rm km}\,{\rm s}^{-1}}\right)^{-1} \left(\frac{\eta_{\rm cr}}{0.1}\right)^{-1}
\] \citep{2010PPCF...52l4006M}, respectively, 
where $V_{\rm a}$ is the Alfv\'en velocity, $L_{\rm cr}$ is the length scale of the CR precursor, $V_{\rm sh}$ is the shock speed, $\eta_{\rm cr}$ is the CR acceleration efficiency, defined by a CR pressure over ram pressure, and $V_{\rm so}$ is the sound velocity.  On the other hand, the crossing time of the precursor is 
\[
t_{\rm cross} \sim 2\times10^{8}\,{\rm s} \left(\frac{L_{\rm cr}}{0.001\,{\rm pc}}\right) \left(\frac{V_{\rm sh}}{200\,{\rm km}\,{\rm s}^{-1}}\right)^{-1}.\]  
A simple comparison of the three time scales reveals that only Alfv\'en waves can have sufficient time to grow. 

The turbulent velocity created by Alfv\'en waves can be expressed as $\delta V$ = $V_{\rm a} \times \frac{\delta B}{B}$, where $B$ and $\delta B$ represent the mean magnetic field and its fluctuation, respectively.  Given that 
\[
\left(\frac{\delta B}{B}\right)^{-2} = \frac{3\kappa}{r_{\rm g}c}
\] and $\kappa = L_{\rm cr} \times V_{\rm sh}$, 
where $\kappa$ is the diffusion coefficient and $r_{\rm g}$ is the gyroradius of CRs, we obtain
\[
\frac{\delta B}{B} \sim 0.4 \left(\frac{L_{\rm cr}}{0.001\,{\rm pc}}\right)^{-0.5} \left(\frac{V_{\rm sh}}{200\,{\rm km}\,s^{-1}}\right)^{-0.5} \left(\frac{B}{3\,\mu{\rm G}}\right)^{-0.5} \left(\frac{p}{m_{\rm p}c}\right)^{0.5}.\]  
Thus, we can estimate 
\[
\delta V \sim 2.7\,{\rm km}\,{\rm s}^{-1} \left(\frac{V_{\rm a}}{6.6\,{\rm km}\,{\rm s}^{-1}}\right) \left(\frac{\frac{\delta B}{B}}{0.4}\right).\]  
If the magnetic field is amplified in the precursor, a somewhat larger turbulent velocity would be expected, as $V_{\rm a}$ linearly depends on $B$ with $\frac{\delta B}{B}$ being expected to stay at unity for strongly amplified magnetic fields.  However, we should keep in mind that the precursor's length scale is inversely proportional to $B$ ($L_{\rm cr} \propto B^{-1}$), and that the precursor's thickness must be larger than the characteristic length of CX (3$\times$10$^{14}$\,cm).  Therefore, the magnetic field would not exceed a few 10 $\mu$G, which results in an upper limit of the turbulent velocity of $\sim$10 km\,s$^{-1}$.  This is significantly smaller than the velocity observed.  Therefore, we conclude that Alfv\'en turbulence can not be the main contributor to the broadening of the narrow component.  

The heating by damping of Alfv\'en waves was extensively investigated in light of observations of the Cygnus Loop by \citet{1988ApJ...333..198B} who revealed that the nonlinear Landau damping \citep{1984A&A...130...19V} can easily heat the upstream gas up to the temperature observed.  In this model, the diffusion coefficient is a key parameter to control the degree of heating.  For the relevant precursor, we can estimate $\kappa \sim 6\times10^{22} \left(\frac{L_{\rm cr}}{0.001\,{\rm pc}}\right) \left(\frac{V_{\rm sh}}{200\,{\rm km}\,{\rm s}^{-1}}\right)$, at which we expect that $T > 2\times10^{6}$\,K according to Fig.~11 in \citet{1988ApJ...333..198B}.  This is much higher than our measurement, $T\sim$3.5$\times10^{4}$\,K.  This means that appropriate fine tuning of the model is required, which is beyond the scope of this $Letter$.  However, we can say that such an excessive heating is probably due to the fact that \citet{1988ApJ...333..198B} assumed that the shock is dominated by CRs, which is unlikely the case for the Cygnus Loop, given the absence of synchrotron X-ray emission, the relatively soft gamma-ray spectrum, and the negligible pressure of nonthermal particles \citep{2009ApJ...702..327S}.

It is interesting to note an order of magnitude difference in the thickness of the CR precursors between the Cygnus Loop and Tycho's Knot g: $L_{\rm cr}\lesssim0.001$\,pc for the Cygnus Loop and $L_{\rm cr}\sim$0.01\,pc for Tycho's Knot g \citep{2010ApJ...715L.146L}.  Combined with different shock speeds of $V_{\rm sh}\sim200$\,km\,s$^{-1}$ \citep[Cygnus Loop:][]{2005AJ....129.2268B} and $\sim3000$\,km\,s$^{-1}$ \citep[Tycho's Knot g:][]{1978ApJ...224..851K,2010ApJ...709.1387K}, an even larger difference is expected for the diffusion coefficient, $\kappa = L_{\rm cr} \times V_{\rm sh}$.  Since $\kappa$ is proportional to $E/B$ with $E$ being the energy of dominant accelerated particles, we speculate that the remarkable difference in $\kappa$ is caused by different energy distribution of accelerated particles between the two sources; there is no doubt that CRs in Tycho's SNR have higher energies than those in the Cygnus Loop, judging from the harder gamma-ray spectrum in Tycho's SNR than that in the Cygnus Loop \citep{2011ApJ...741...44K,2012ApJ...744L...2G}.  No firm conclusion can be reached yet, however, without information about the value of $B$.  Whatever the nature, the difference of thin precursors' sizes between the Cygnus Loop and Tycho's Knot g gives us a strong motivation to search for a positive correlation between the size of a thin precursor (or diffusion coefficient) and the hardness of CR energy distribution.


\acknowledgments

We thank all the members of the {\it Subaru} telescope, especially Akito Tajitsu, for performing our observation.  We also thank Masaomi Tanaka for helping our proposal.  This work is supported by Japan Society for the Promotion of Science KAKENHI Grant Numbers 25800119 (SK), 26800100 (KM), and 15K05088 (RY).  The work by KM is partly supported by World Premier International Research Center Initiative (WPI Initiative), MEXT, Japan.  JCR's work was supported by grant HST-GO-13436 to the Smithsonian astrophysical Observatory.




\end{document}